# Fast and Rigorous DC Solution in Finite Element Method for Integrated Circuit Analysis

Q. He

Purdue University

465 Northwestern Ave., West Lafayette IN 47907

Large scale circuit simulation, such as power delivery network analysis, has become increasingly challenge in the VLSI design verification flow. Power delivery network can be simulated by both SPICE-type circuit-based model and eletromagnetics-based model when full-wave accuracy is desired [7, 8]. In the early time of the time domain finite element simulation [1] for integrated circuit, the modes having the highest eigenvalues supported by the numerical system will be excited [2-5]. Because of the band limited source, after the early time, the modes having a resonance frequency well beyond the input frequency band will die down, and all physically important high-order modes and DC mode will show up and become dominant [9-12]. Among these modes, the DC mode is the last one to show up. Although the convergence criterion is not applied on the DC mode, the existence of DC mode in the field solution will deteriorate the convergence rate of the first several high order modes. Therefore, this paper first analyzed the mathematic characteristics of the DC mode and proposed a rigorous and fast solution to extract the DC mode from the numerical system in order to speed up the convergence rate. Experimental results demonstrated the robustness and superior performance of this method.

## 1. Theoretical Analysis of the DC Mode in a Coupled System

Without loss of generality, this theoretical analysis is demonstrated in a lossless finite element system constructed by two triangle elements as illustrated in Fig. 1 (a).

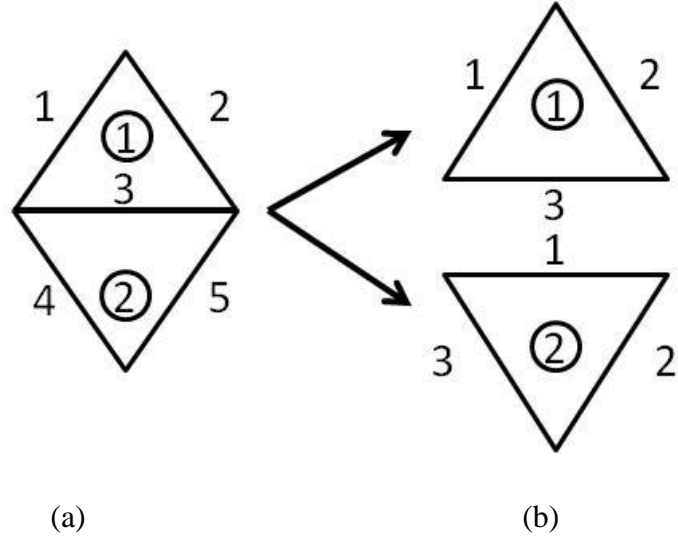

(a)                              (b)

Fig. 1 A Two-triangle finite element system: (a) Coupled system with global edge number; (b) Two standalone sub-systems with local edge number

An FEM based matrix equation is [1]

$$(\mathbf{S} - \omega^2 \mathbf{T})x = b(\omega) \ . \quad (1)$$

The solution of (1) can be obtained by solving the following generalized eigenvalue problem that is frequency independent [12]

$$\mathbf{S}v = \lambda \mathbf{T}v , \quad (2)$$

where $\lambda$ is the eigenvalue, and $v$ is the eigenvector. Since S is symmetric semi-positive definite and T is symmetric positive definite, the eigenvalues are non-negative real numbers. Meanwhile, we have the properties

$$\mathbf{V}^T\mathbf{T}\mathbf{V}=\mathbf{I}, \ \mathbf{V}^T\mathbf{S}\mathbf{V}=\mathbf{\Lambda}, \quad (3)$$

where $\mathbf{V}=[v_1, v_2,... v_n]$, $\mathbf{I}$ is an identity matrix, and $\mathbf{\Lambda}$ is a diagonal matrix, the $i$th entry of which is $\lambda_i$. The field solution $x$ can be obtained by

$$x=\mathbf{V}y, \quad (4)$$

where $y$ is the coefficient vector associated with the eigenvectors. If the field solution only has DC mode, $x$ is constructed by

$$x=\mathbf{V}_0 y_0, \quad (5)$$

where null space $\mathbf{V_0}$ denotes the eigenvectors corresponding to zero eigenvalues. $\mathbf{V_0}$ satisfy

$$\mathbf{SV_0} = 0, \tag{6}$$

which can be seen from (2). Each row of the matrix-vector multiplication $\mathbf{SV_0}$ is an assembled

$$\mu_r^{-1} \langle \nabla \times \mathbf{N}_i, \nabla \times \mathbf{E} \rangle, \tag{7}$$

where $\mathbf{N}_i$ is the *i*-th vector basis. The nonzero solution of (6) must satisfy $\nabla \times \mathbf{E} = 0$, and thereby being a gradient field.

Here, we realize an important property of DC mode that the null space $\mathbf{V_0}$ constructed from each standalone sub-domain, as illustrated in Fig 1(b), provide the right space of the sub-domain in the coupled system. Hence, instead of solving the global large eigenvalue problem, we can solve the eigenvalue problem in each small sub-domain. The rigorous proof of this property is as follows.

Consider the coupled system in Fig. 1(a), from (6) we have

$$\begin{bmatrix} \mathbf{S}_{11} & \mathbf{S}_{12} \\ \mathbf{S}_{21} & \mathbf{S}_{22} \end{bmatrix} \begin{Bmatrix} X_1 \\ X_2 \end{Bmatrix} = 0, \tag{8}$$

where $X_1$ and $X_2$ are the field solution in region 1 and 2. The second row of (8) is

$$\mathbf{S}_{21} X_1 + \mathbf{S}_{22} X_2 = 0, \tag{9}$$

where the *i*th row of (9) is formed by

$$\begin{aligned} &\mu_r^{-1} \langle \nabla \times \mathbf{N}_i, \nabla \times \mathbf{N}_1^1 \rangle x_1 + \mu_r^{-1} \langle \nabla \times \mathbf{N}_i, \nabla \times \mathbf{N}_2^1 \rangle x_2 + \\ &\mu_r^{-1} \left( \langle \nabla \times \mathbf{N}_i, \nabla \times \mathbf{N}_3^1 \rangle + \langle \nabla \times \mathbf{N}_i, \nabla \times \mathbf{N}_1^2 \rangle \right) x_3 + \\ &\mu_r^{-1} \langle \nabla \times \mathbf{N}_i, \nabla \times \mathbf{N}_3^2 \rangle x_4 + \mu_r^{-1} \langle \nabla \times \mathbf{N}_i, \nabla \times \mathbf{N}_2^2 \rangle x_5 = 0 \end{aligned} \tag{10}$$

in which, the superscript of the basis function $\mathbf{N}$ denotes the element number, and the subscript denotes the local edge in this element. Reorder (10), we have

$$\begin{aligned} &\mu_r^{-1} \langle \nabla \times \mathbf{N}_i, \nabla \times (\mathbf{N}_1^1 x_1 + \mathbf{N}_2^1 x_2 + \mathbf{N}_3^1 x_3) \rangle + \\ &\mu_r^{-1} \langle \nabla \times \mathbf{N}_i, \nabla \times (\mathbf{N}_1^2 x_3 + \mathbf{N}_3^2 x_4 + \mathbf{N}_2^2 x_5) \rangle = 0 \end{aligned}, \tag{11}$$

which can be written as

$$\mu_r^{-1} \langle \nabla \times \mathbf{N}_i, \nabla \times \mathbf{E}_1 \rangle + \mu_r^{-1} \langle \nabla \times \mathbf{N}_i, \nabla \times \mathbf{E}_2 \rangle = 0. \tag{12}$$

From (12) we can see that

$$\nabla \times \mathbf{E}_1 = 0 \text{ and } \nabla \times \mathbf{E}_2 = 0 \tag{13}$$

is a set of solution of (12), or we can say that the nullspace in each standalone sub-region which satisfy

$$\mathbf{S}_1^0 \mathbf{V}_0^1 = 0 \text{ and } \mathbf{S}_2^0 \mathbf{V}_0^2 = 0, \tag{14}$$

where $\mathbf{S}_1^0$ and $\mathbf{S}_2^0$ are the stiffness matrix of the standalone sub-region 1 and 2 in Fig. 1, and $\mathbf{V}_0^1$ and $\mathbf{V}_0^2$ denote the nullspace of $\mathbf{S}_1^0$ and $\mathbf{S}_2^0$ respectively, provide the nullspace of the DC field distribution in the sub-region of the global large problem which satisfies (6). Therefore, the nullspace is obtained by the standalone sub-region stiffness matrix $\mathbf{S}$ only. Since $\mathbf{S}$ is a real valued matrix, it is more efficient to obtain the eigenvalue solution of S than that of the complex values matrix $\mathbf{S}+j\omega\mathbf{R}-\omega^2\mathbf{T}$. Another benefit of $\mathbf{S}$'s eigenvalue analysis is that since $\mathbf{S}$ is only related to the mesh regardless of material parameters, the $\mathbf{S}$'s eigenvalue analysis does not need to repeat for regions having different materials as long as the mesh is the same.

## 2. A Rigorous Divide and Conquer Approach to Extract the Global DC Mode

It is clear to use the divide and conquer methodology to break the original large-scale global problem down to each sub-domain. However, it does not mean that we can solve (1) in the standalone sub-region to find field solution in each small domain instead of the global problem of (1), because all sub-regions are coupled together at the interface. The nullspace $\mathbf{V}_0^i$ in each standalone region provides the basis to generate the field solution in each domain, which is the superposition of each null space vector multiplying a corresponding coefficient:

$$x_i = \mathbf{V}_0^i y_i. \tag{15}$$

The right coupling information from other regions are calculated in the coefficient vector $y_i$.

A straightforward approach to solve (1) by (15) is to build the global nullspace $\mathbf{V}_0$ in (5) by $\mathbf{V}_0^i$ from the eigenvalue solution of $m$ standalone sub-regions' stiffness matrix $\mathbf{S}_i^0$, where

$$\mathbf{V}_0 = \begin{bmatrix} \mathbf{V}_0^1 & & & \\ & \mathbf{V}_0^2 & & \\ & & \ddots & \\ & & & \mathbf{V}_0^m \end{bmatrix}. \tag{17}$$

Substituting (16) into (5) and then (5) into (1), multiplying both sides of (1) by $\mathbf{V}_0^T$, we obtain

$$y_0 = \left[\mathbf{V}_0^T (\mathbf{S} - \omega^2 \mathbf{T})\mathbf{V}_0\right]^{-1} \mathbf{V}_0^T b(\omega). \tag{17}$$

Here we need to point out that this $\mathbf{V}_0$ from (16) does not have the property $\mathbf{V}_0^T \mathbf{S} \mathbf{V}_0 = 0$ because $\mathbf{V}_0^i$ only makes $\mathbf{S}_i^0$ equal to zero not the $\mathbf{S}_{ii}$ part in $\mathbf{S}$ equal to zero. Substituting (17) into (5), we have the field solution at DC

$$x = \mathbf{V}_0 \left[\mathbf{V}_0^T (\mathbf{S} - \omega^2 \mathbf{T})\mathbf{V}_0\right]^{-1} \mathbf{V}_0^T b(\omega). \tag{18}$$

The efficiency in this approach is not optimal because $\mathbf{V}_0^T(\mathbf{S}-\omega^2\mathbf{T})\mathbf{V}_0$ is not an identity matrix. The complexity of this inverse is $O(N^3)$ which is the same as the complexity of solving (1). We can project the Schur complement of region $i$ onto $\mathbf{V}_{0,i}$ and solve the problem in each sub-region recursively to avoid the expensive inverse computation in (18). Here we still use the problem illustrated in Fig. 1 as the example.

(1) can be re-written as

$$\begin{bmatrix} \mathbf{A}_{11} & \mathbf{A}_{12} \\ \mathbf{A}_{21} & \mathbf{A}_{22} \end{bmatrix} \begin{Bmatrix} x_1 \\ x_2 \end{Bmatrix} = \begin{Bmatrix} b_1 \\ b_2 \end{Bmatrix}, \tag{19}$$

where $\mathbf{A}_{11}=\mathbf{S}_{11}-\omega^2\mathbf{T}_{11}$, $\mathbf{A}_{12}=\mathbf{S}_{12}-\omega^2\mathbf{T}_{12}$, $\mathbf{A}_{21}=\mathbf{S}_{21}-\omega^2\mathbf{T}_{21}$, and $\mathbf{A}_{22}=\mathbf{S}_{22}-\omega^2\mathbf{T}_{22}$. $x_2$ can be solved by the Schur complement $\tilde{\mathbf{A}}_{22}$ as

$$\tilde{\mathbf{A}}_{22} x_2 = b_2', \tag{20}$$

where

$$\tilde{\mathbf{A}}_{22} = \mathbf{A}_{22} - \mathbf{A}_{21}\mathbf{A}_{11}^{-1}\mathbf{A}_{12} \tag{21}$$

and

$$b_2' = b_2 - \mathbf{A}_{21}\mathbf{A}_{11}^{-1}b_1. \tag{22}$$

Since $\mathbf{V}_{0,2}$ is the space of $x_2$, substituting

$$x_2 = \mathbf{V}_{0,2} y_2 \tag{23}$$

into (20) and multiplying both sides of (20) with $\mathbf{V}_{0,2}^T$, we obtain

$$y_2 = \left[ \mathbf{V}_{0,2}^T \tilde{\mathbf{A}}_{22} \mathbf{V}_{0,2} \right]^{-1} \mathbf{V}_{0,2}^T b_2'(\omega). \tag{24}$$

Substituting (24) into (23), we have the field solution in region 2 as

$$x_2 = \mathbf{V}_{0,2} \left[ \mathbf{V}_{0,2}^T \tilde{\mathbf{A}}_{22} \mathbf{V}_{0,2} \right]^{-1} \mathbf{V}_{0,2}^T b_2'(\omega). \tag{25}$$

Compared with (18), the size of the inverse in (25) is only half of the one in (18) if the two regions are equally partitioned. We can partition the system to each layer to calculate the solution recursively with optimal complexity. The nullspace size in each layer is much smaller than the nullspace size of the entire system.

## 3. Numerical Result of the Fast DC Extraction Method

In this section, we provide four examples to valide the proposed method. The first example is a lossless parallel plate. The second is a bus structure. The third is an on-chip power grid example. The last one a realistic single stripline structure provided by Intel. In all these examples, we compare our solution with the solution from the global eigenvalue method and the matrix inverse solution to demonstrate the accuracy of the proposed method.

### 3.1 Lossless Parallel Plate Waveguide

According to typical on-chip circuit dimensions, the waveguide width (along $y$) is set as 4 μm, the height (along $x$) is set as 1 μm, and the waveguide length (along $z$) is set as 100 μm. The spatial discretization is chosen as $\Delta x = 0.33$ μm, $\Delta y = 1$ μm, and $\Delta z = 20$ μm. The structure is discretized by the brick finite element. A current source is launched at the near end of the waveguide from bottom PEC plane to top PEC plane. The frequency point was chosen at 2 GHz, which is low enough to excite DC field only and not too low to avoid breakdown. The upper bound of the frequency choice is determined by the largest physical dimension of the structure since the first resonance frequency is calculated from the largest physical dimension. The lower bound of the frequency range is determined by the smallest mesh size because the largest eigenvalue in the numerical system is determined by the smallest mesh size.

The reference solutions are obtained by both the global eigenvalue solution of (7-8) and the matrix inverse solution of (3). Total number of unknowns is 188. The global eigenvalue solution has a nullspace of size 61.

In the proposed method, the computational domain is divided to 5 sub-domains in z direction with one layer in each domain. The first layer is composed of one standalone surface, one coupled surface, and one volume, denoted by $s^0$-v-s. All other layers consist of one coupled surface and one volume, denoted by v-s, except for the last layer which is formed by one standalone surface and one volume. This partition is illustrated in Fig. 2.

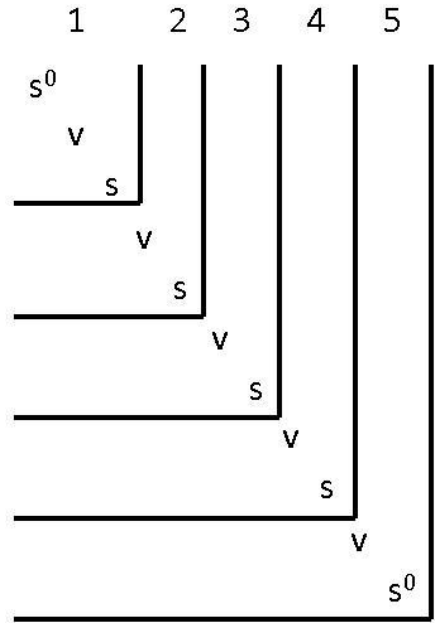

Fig. 2 Matrix partition of the parallel plate example

The standalone $S_{ii}^0$ of one layer is formed by unknowns on two standalone-surfaces and unknowns in one volume, denoted by $s^0$-v-$s^0$. The dimension of $S_{ii}^0$ is 56. After the eigenvalue analysis of $S_{ii}^0$, we find 21 zero eigenvalues. Therefore, the nullspace is constructed by 21 eigenvectors associated with the zero eigenvalues. The zero eigenvalues numerically are not zeros. But they can be easily identified in the eigenvalue distribution because of the big gap between the zero eigenvalues and the high-order eigenvalues. Table 1 shows all eigenvalues found in the eigenvalue analysis of $S_{ii}^0$.

We select 33 rows of the nullspace which belong to the surface and volume unknowns in this sub-domain to build $\mathbf{V}_0$ with size 33 by 21. The size of $\mathbf{B}_i$ is 21 by 21.

Table 1 Eigenvalue distribution of the standalone $\mathbf{S}_{ii}^0$ of one layer in the parallel plate waveguide

|    | DC        | High order |
|----|-----------|------------|
| 1  | -1.44E-20 | 1.59E-07   |
| 2  | -1.01E-20 | 51E-07     |
| 3  | -14E-21   | 1.91E-06   |
| 4  | -6.59E-21 | 2.41E-06   |
| 5  | -4.63E-21 | 4.45E-06   |
| 6  | -3.53E-21 | 6.10E-06   |
| 7  | -2.62E-21 | 6.91E-06   |
| 8  | -1.62E-21 | 46E-06     |
| 9  | -1.43E-21 | 9.09E-06   |
| 10 | 4.60E-22  | 1.11E-05   |
| 11 | 8.96E-22  | 1.28E-05   |
| 12 | 1.23E-21  | 1.31E-05   |
| 13 | 1.96E-21  | 1.81E-05   |
| 14 | 2.13E-21  | 2.07E-05   |
| 15 | 3.68E-21  | 2.17E-05   |
| 16 | 4.20E-21  | 2.17E-05   |
| 17 | 5.28E-21  | 2.19E-05   |
| 18 | 79E-21    | 2.24E-05   |
| 19 | 1.04E-20  | 2.44E-05   |
| 20 | 1.17E-20  | 2.61E-05   |
| 21 | 1.29E-20  | 2.69E-05   |
| 22 |           | 3.31E-05   |
| 23 |           | 3.81E-05   |

| 24 | | 3.97E-05 |
|---|---|---|
| 25 | | 4.05E-05 |
| 26 | | 4.12E-05 |
| 27 | | 4.36E-05 |
| 28 | | 4.59E-05 |
| 29 | | 6.19E-05 |
| 30 | | 6.69E-05 |
| 31 | | 26E-05 |
| 32 | | 31E-05 |
| 33 | | 81E-05 |
| 34 | | 9.47E-05 |
| 35 | | 1.13E-04 |

The accuracy of this method is validated by comparing the solution vector from the proposed method with the reference results. Reference 1 is the global eigenvalue solution. Reference 2 is the matrix inverse solution. From Table 3, good accuracy was observed. We also verified the result with the analytical solution since at DC the parallel plate is purely a capacitance and the capacitance value can be analytically obtained by $C = \varepsilon \frac{A}{d}$, where $\varepsilon$ is the permittivity of the material between the plates, $A$ is the area of overlap of the two plates, and $d$ is the separation between the plates. From our result, we know the current $I$ and voltage $V$ in this parallel plate. The capacitance is calculated by $C = \frac{I}{j\omega V}$. Tow capacitance values agree well with each other.

Table 2 Relative error of the proposed method with the reference in the parallel plate example

| $\frac{|x_{prop} - x_{ref1}|}{|x_{ref1}|}$ | 0.000373522201289469 |
|---|---|

| $\dfrac{|x_{prop} - x_{ref2}|}{|x_{ref2}|}$ | 0.000373534937419577 |

### 1.1.1 Lossy Two Bus Structure

The second example is a lossy two-bus structure. The structure and discretization are shown in Fig. 3. The bus is 500 μm long truncated by air in front and back, PEC on top and bottom, PMC on left and right. The dielectric constant is 4.1. The two buses are made of copper. A current source at 10 GHz is launched between two buses. The structure has 12 layers in z direction. Each layer is a sub-domain. The nullspace of standalone $S_{ii}^0$ in one layer has 91 independent vectors. From the error results shown in Table 3, we can observe very good agreement of the proposed method with the reference solution.

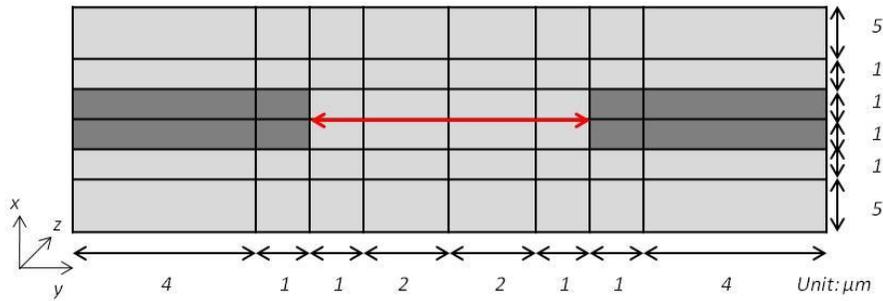

Fig. 3 Lossy two-bus structure size and discretization. The dark color is the metal bus

Table 3 Relative error of the proposed method with the reference in the on-chip bus example

| $\dfrac{|x_{prop} - x_{ref1}|}{|x_{ref1}|}$ | 3.6e-6 |
| $\dfrac{|x_{prop} - x_{ref2}|}{|x_{ref2}|}$ | 3.6e-6 |

## 3. Conclusion

In this paper, a fast DC mode extraction algorithm is developed by decomposing the original large-scale problem rigorously into small problems that are fully decoupled, and then synthesizing the DC solution of the original large-scale problem from the nullspace of the small problems. This fast DC mode extraction method has a lot of applications. For example, by using this fast DC mode extraction method, we can significantly speed up the premarching process in the explicit time domain unconditionally stable method [2, 4] by removing the DC field solution from the premarching part since the DC mode is the slowest one to converge. As a result, the unconditionally stable method is further accelerated because the premarching time is shortened significantly. Numerical examples have demonstrated the accuracy and efficiency of the proposed method.